\documentclass{amsproc}

 \usepackage{setspace}
\usepackage[all]{xypic}
\usepackage{scrlayer-scrpage}
 \usepackage{enumerate}
\usepackage{multirow}
\usepackage{amsmath}
\usepackage{mathtools}
\usepackage{mathrsfs}

\usepackage{amssymb}
\usepackage[square,numbers]{natbib}
\usepackage{tikz,pgf}
\newtheorem{theorem}{Theorem}[section]

\theoremstyle{definition}

\theoremstyle{remark}

\renewcommand{\tilde}{\widetilde}

\newcommand{\fsu}{\mathfrak{su}}
\newcommand{\fso}{\mathfrak{so}}
\newcommand{\fsp}{\mathfrak{sp}}

\newcommand{\ff}{\mathfrak{f}}
\newcommand{\fe}{\mathfrak{e}}

\newcommand\be{\begin{equation}}
\newcommand\ee{\end{equation}}
\renewcommand{\hat}{\widehat}

\newcommand\SEITcc[9]
{$\xymatrix@=0.01pc{&&&\tempi \atop \qquad\\
&&&\overset{\mathfrak{\fsp}_{#8}}{1}\\ &\tempa\,\,\overset{\mathfrak{\fso}_{#1}}{4}&\underset{\tempb}{\overset{\mathfrak{\fsp}_{#2}}{1}}&\underset{\tempc}{\overset{\mathfrak{\fso}_{#3}}{4}}&\underset{\tempd}{\overset{\mathfrak{\fsp}_{#4}}{1}}&\underset{\tempe}{\overset{\mathfrak{\fso}_{#5}}{4}}&\underset{\tempf}{\overset{\mathfrak{\fsp}_{#6}}{1}}&\underset{\tempg}{\overset{\mathfrak{\fso}_{#7}}{4}}&\overset{\mathfrak{\fsp}_{#9}}{1}\tempj}$}

\numberwithin{equation}{section}

\begin{document}

\title[6D Heterotic LST]{6D Heterotic Little String Theories and F-theory Geometry: An Introduction}
%\title{}


%    author one information
\author[M. Del Zotto]{Michele Del Zotto}
%\author{}
\address{Department of Mathematics, Uppsala University, Uppsala, Sweden}
\address{Department of Physics and Astronomy, Uppsala Univeristy, Uppsala, Sweden}
\curraddr{}
\email{michele.delzotto@math.uu.se}
\thanks{}

%    author two information
\author[M. Liu]{Muyang Liu}
\address{Department of Mathematics, Uppsala University, Uppsala, Sweden}
\curraddr{}
\email{muyang.liu@math.uu.se}
\thanks{}

%    author three information
\author[P.-K. Oehlmann]{Paul-Konstantin Oehlmann}
\address{Department of Mathematics, Uppsala University, Uppsala, Sweden}
\curraddr{Department of Physics, Northeastern University, Boston, USA}
\email{p.oehlmann@northeastern.edu}
\thanks{}

%    The 2020 edition of the Mathematics Subject Classification is
%    the current definitive version.
\subjclass[2020]{Primary }

\date{}

\begin{abstract}
We review here some aspects of our recent works about the geometric engineering of heterotic little string theories using F-theory. Building on the seminal work by Aspinwall and Morrison as well as Intrilligator and Blum, we solve some longstanding open questions thanks to recent progress in our understanding of 6D (1,0) theories and their generalized symmetries. On the geometry side, these systems correspond to non-compact elliptically fibered Calabi-Yau varieties that must admit the structure of an elliptic K3 fibration. From fiberwise F-theory/Heterotic duality the K3 plays a central role - it determines the 6D flavor group, as well as different T-dual LSTs via inequivalent elliptic fibration structures. 
The geometries we obtain are some finer versions of Kulikov degenerations: the point where the K3 fiber degenerates is the locus where the LST arises. This structure serve on one hand to check our field theory predictions on LST dualities via the match of Coulomb branch dimension, flavor symmetries, and 2-group structure constants, and also on the other hand to deduce novel LST models and their networks of dualities, thus allowing exploring non-geometric Heterotic regimes.

\end{abstract}

\maketitle

\section{Introduction and Summary}

Six dimensional supersymmetric theories can be divided into three classes: Supergravity theories (SUGRAs), little string theories (LSTs), and superconformal field theories (SCFTs). Of these, the latter two are distinguished as they are decoupled from 6D gravity. There is typically a hierarchical chain of 6D decoupling limits: SUGRAs come with a corresponding 6D Planck scale $M_{PL}$, which controls the strenght of gravitational interactions, LSTs similarly comes with an intrinsic string tension $M_{LS}$ (much smaller than $M_{PL}$), while SCFTs have no scales whatsoever. In 6D, lowering the energy, one typically flows down from a SUGRA to a LST and then eventually to a SCFT. The LST string tension is an energy scale that breaks conformal invariance, at the price of allowing features usually found in string theories. In the latter, the Planck scale can mix with the Kaluza-Klein scale upon circle reduction which realizes T-duality (as found eg. in the $E_8\times E_8 \leftrightarrow Spin(32)/\mathbb{Z}_2$ heterotic duality). A similar mechanism applies in LSTs, that can be T-dual upon circle reduction since the KK scale can mix with the LST string tension exchanging winding and momentum modes. This signals these systems are not conventional quantum field theories, as T-duality indicates these systems have multiple stress energy tensors. Decoupling the little strings, considering energies well below $M_{LS}$, one ends up with an SCFT in the infrared
\cite{Seiberg:1997zk,Intriligator:1997dh, Intriligator:1999cn,Hanany:1997gh,Brunner:1997gf,Kapustin:1999ci}
.

Due to the advent of F-theory, powerful geometric methods became available to substantially improve our understanding of 6D systems. The hierarchy we outlined above is reflected by F-theory geometry: for SUGRA the 6D planck scale is set by the volume of the F-theory base, sending the latter to infinity the 6D graviton decouples. This decoupling limit is equivalent to focusing on a shrinking surface: the latter can either shrink to a curve (the LST case: the volume of such curve sets the scale $M_{LS}$) or to a point (the SCFT case). The resulting geometries are so constrained that it leads to a classification  \cite{Heckman:2015bfa,Bhardwaj:2015oru}.\footnote{ See eg. \cite{Heckman:2018jxk} or Section 2 of \cite{DelZotto:2018tcj} for a review and further references to recent works on the subject.} This approach allowed to re-cast the physics of LSTs and SCFTs in geometrical properties of non-compact elliptic Calabi-Yau threefolds $X_3$ exploiting standard F-theory techniques \cite{Vafa:1996xn,Morrison:1996na,Morrison:1996pp,
Morrison:1996xf,Aspinwall:1997ye,Katz:2011qp} (as well as stimulating the development of some novel featurs of F-theory \cite{Tachikawa:2015wka,Bhardwaj:2018jgp}). 

In this review we summarise some of the results in our papers \cite{DelZotto:2022ohj,DelZotto:2022xrh,DZLOP3} where we are interested in a special class of 6D LSTs which arises as the theory governing the worldvolumes of NS5 branes in Hetertotic string theories in presence of an ALE singularity of type $\mathbb C^2/\Gamma$ where $\Gamma$ is a discrete subgroup of $SU(2)$. For the $Spin(32)/\mathbb Z_2$ cases these models have been completely determined by \cite{Blum:1997mm,Intriligator:1997dh} --- since the theory of NS5 branes of the $Spin(32)/\mathbb Z_2$ is Lagrangian \cite{Witten:1995gx}, the theories corresponding to the worldvolume ALE instantons are orbifolds of the latter which depend on the embedding of $\Gamma$ into $Spin(32)/\mathbb Z_2$ (corresponding to a choice of a flat connection at infinity for the Heterotic bundle). For the $E_8\times E_8$ theories this is not the case, however: the theories for the Heterotic $E_8\times E_8$ instantons do not have a conventional Lagrangian description \cite{Ganor:1996mu}, and therefore the corresponding ALE instantons are much trickier to determine \cite{Aspinwall:1997ye}. Thanks to recent advances in our understanding of 6D SCFTs, however, one can now characterize them in terms of generalized 6D quivers, exploiting fusion of specific 6D building blocks \cite{DelZotto:2014hpa,Heckman:2018pqx}. The 6D SCFT building blocks of interest for us are 6D conformal matter models \cite{DelZotto:2014hpa} and orbi-instanton 6D SCFTs \cite{Mekareeya:2017jgc,Heckman:2018pqx,Frey:2018vpw}, which have been completely classified. This strategy gives a full characterization of the Heterotic $E_8\times E_8$ ALE instantons models and their dependence on choices of $E_8\times E_8$ flat connections at infinity in terms of generalized 6D quiver diagrams. Each quiver diagram encodes also the structure of an ellipic fibration for the corresponding F-theory geometry.

As a consistency check for this result one should still verify that for each such model there is a corresponding T-dual $Spin(32)/\mathbb Z_2$ theory. In order to constrain the matching duals we exploit the fact that, as opposed to gravitational theories, LSTs can have global symmetries,\footnote{
In gravity theories it is widely believed that all symmetries must either be broken or gauged which also applies to their generalized versions (see \cite{Banks:2010zn} for a general argument, and \cite{Apruzzi:2020zot,Braun:2019wnj} for a sample of applications to 6D).} which must match across dualities. In particular, a stringent constraint about the T-duality of LSTs is given by the matching of their 2-group symmetries \cite{DelZotto:2020esg}. 2-groups symmetries are examples of generalized global symmetries \cite{Gaiotto:2014kfa} where a symmetry which acts on point operators mixes with a symmetry acting on line operators.\footnote{For an incomplete list of references about higher symmetries of 6D systems see eg. \cite{DelZotto:2015isa,Bhardwaj:2020phs,Apruzzi:2021mlh,
Hubner:2022kxr,Cvetic:2022imb}.}  LSTs in particular always have a universal continuous 2-group symmetry which couples the 2-form background field corresponding to the little string charge with the 1-form background fields for the Poincaré symmetry, the R-symmetry and the other global symmetries of the model \cite{Cordova:2018cvg,Cordova:2020tij,DelZotto:2020esg}. Schematically the LST 2-group is presented as follows
 \begin{align}
^2 G_{ \hat \kappa_\mathscr{P}, \hat \kappa_\mathscr{R}, \hat \kappa_{\mathscr{F}_A}} =\left( 
\mathscr{P}^{(0)} \times \mathscr{R}^{(0)}\times \mathscr{F}_A^{(0)} \right) \times_{\hat \kappa_\mathscr{P},  \hat \kappa_\mathscr{R},\hat \kappa_{\mathscr{F}_A}
} U(1)^{(1)}_{LS} \, ,
 \end{align} 
where $U(1)^{(1)}_{LS}$ is the 1-form symmetry measuring the little string charge (whose background field is the 2-form above), $\mathscr{P}$ is the spacetime Poincar\'e symmetry, $\mathscr{R}^{(0)} \simeq \text{SU}(2)_{R}$ is the R-symmetry of the (1,0) supersymmetry, and various other flavor 0-form symmetries  are denoted $\mathscr{F}_A^{(0)}$. Each component of this 2-group is characterized by a 2-group structure constant denoted $\hat \kappa$ above which measures the "mixing" between these symmetries \cite{Cordova:2018cvg,Cordova:2020tij,DelZotto:2020esg}. In particular, $\hat \kappa_{\mathscr P}$ can have only two values $0$ or $2$. In the former case, Poincar\'e symmetry does not mix with $U^{(1)}_{LS}$, in the latter it does. Remarkably, this coincides with the number of M9 branes in an M-theory dual formulation.\footnote{Recall that there is a dual description of the $E_8 \times E_8$ Heterotic in M-theory, where M-theory is placed on $S^1/\mathbb Z_2$ and two M9 branes are placed at the fixed points. This is the so called Hořava-Witten (HW) description of the Heterotic \cite{Horava:1995qa,Horava:1996ma}.} It is easy to check that, as expected from the HW description, for the Heterotic LSTs $\hat \kappa_{\mathscr P}$ always equals 2. On the contrary $\hat \kappa_\mathscr{R}$ has an interesting structure and its matching is much more constraining \cite{DelZotto:2020sop}. The 2-groups provide a powerful novel invariant to explore the network of T-dual theories and their relations, also for the case of Heterotic ALE instanton theories. This gives a strategy to conjecture T-dual pairs of little string ALE instantons on the $E_8\times E_8$ and the $Spin(32)/\mathbb Z_2$ Heterotic theories. The geometry of  F-theory gives the most powerful tool to verify these predictions.

 Proving T-duality explicitly can be achieved very easily when a perturbative brane picture exists \cite{Blum:1997mm,Intriligator:1997dh,Hanany:1997gh,Brunner:1997gf}. However this is not the case for exceptional singularities, which do not admit a description in perturbative string theory. In order to tackle those one must exploit F-theory geometry. in facts, F-theory allows also to reproduce (independently) the results from perturbative brane setups, and it allows to consider the exceptional cases on the same footing, thus giving a nice unifying perspective. In F-theory the type IIB axio-dilation is 
realized as the torus modulus of an auxiliary elliptic fiber over the physical base space, which completes the total space to an elliptic Calabi-Yau (CY) threefold \cite{Vafa:1996xn}. The F-theory framework allows to prove T-dualities geometrically through its beautiful connection to M-theory \cite{Aspinwall:1996vc,Aspinwall:1997ye,Bhardwaj:2015oru}: 
M-theory on an elliptic threefold $X_3$ is dual to  F-theory on the same CY times a circle. The radius of the circle in F-theory is inversely proportional to the volume of the elliptic fiber. Sending the radius to infinity, getting back to 6D, the elliptic fiber shrinks to zero size. The M-theory moduli space is therefore larger than the F-theory moduli space. In particular, the circle reduction introduces Wilson lines for the 6D gauge fields that become  Coulomb branch parameters of  the 5D theory which add to the scalars from the 6D tensorbranch to give rise to a larger 5D Coulomb branch. Those parameters are identified with the K\"ahler moduli of the underlying threefold that allows to resolve the singularities in the elliptic fiber (which one cannot do in 6D). A well-known feature of 5D compactifications is that the Coulomb branch splits into chambers, separated by walls where some BPS particles become massless \cite{Seiberg:1996bd}. In geometry, these give flop transitions which lead to different \textit{birational equivalent} CYs being part of the same moduli space \cite{Morrison:1996xf}. Toghether these form the so-called extended K\"ahler cone. With M-theory one can probe that whole space, which is not accessible in F-theory.

Two 6D LSTs are T-dual whenever they give rise to the same 5D theory upon circle compactification.  In the geometry this is realized from the F-theory/M-theory duality above as follows. The elliptic CY giving rise to the two LSTs in question have two inequivalent elliptic structures which however belong to the same M-theory moduli space. The resulting geometry then must have two inequivalent ways to be lifted up to F-theory, thus leading a 5D theory that has two inequivalent 6D origins. This is the hallmark of T-duality. It can happen that a single CY admits two inequivalent elliptic fibrations or that the two elliptically fibered CYs are birational equivalent \cite{Aspinwall:1996vc}: in either case, the two belong to the same M-theory moduli space, while being disconnected in the F-theory moduli space \cite{Bhardwaj:2015oru}. Moreover, it can happen that the given CY admits more than just two inequivalent elliptic structures, thus giving raise to multiple 6D LSTs becoming equivalent theories in 5D \cite{DelZotto:2022ohj,Bhardwaj:2022ekc}.\footnote{Notice that since these models are equivalent 5D theories the resulting 5D flavor symmetries must be identical. This often requires turning on a flavor Wilson line on the T-duality circle, which breaks the 6D flavor symmetries down to a common subgroup. In particular, the flavor group ranks must match, which is built in geometry: denoting $d_{CB}$ the 5D Coulomb branch dimension and $f$ the rank of the flavor symmetry $\text{dim}H_2(X_3) = d_{CB} + f$, while $\text{dim}H_4(X_3) = r_{5D}$. The Betti numbers match for birational equivalent 3-folds.}

The above approach gives a geometrical way to establish T-dualities and it allows to search for novel duals systematically. In the series of work \cite{DelZotto:2022ohj,DelZotto:2022xrh,DZLOP3}  we have focused on heterotic LSTs for which we have identified a rich geometric substructure. This structure allows 
 for general statements about this class of LSTs as well as potential new directions in mathematics which we summarize as follows:
\begin{enumerate}
   \item \textbf{Universal Elliptic K3 structure:}
  Each heterotic LST, admits a geometric engineering approach where  the non-compact elliptic threefold $X_3$ admits a nested elliptic K3 fibration structure.  
 \item\textbf{Flavor from the K3:} The K3 contains non-compact fibral divisors which yield the flavor group of the LST. For a fixed K3 polarization, it is given via the frame lattice $W$. This establishes a correspondence between heterotic instantons labeled by flat connections at infinity and degenerations of K3s with a given frame lattice, \textit{à la}  Friedman-Morgan-Witten \cite{Friedman:1997yq}
 \item\textbf{LST Dualities from K3:}
 For a fixed $X_3$, all inequivalent elliptic fibrations must descend from the flavor K3 fiber. This allows to classify all T-dual LSTs corresponding to $X_3$ from the flavor K3 alone via lattice techniques.\footnote{The extended Kähler cone has the potential for a much richer landscape of interconnections: one can have examples where $X_3$ is birational to a different elliptic CY which has an inequivalent nested K3 fibration leading to another family of T-dual Heterotic theories. Remarkably however, in all the examples we have considered so far, a single nested K3 fibration seems to suffice to determine all the (non twisted) T-duals.\label{WTF}}
 \item \textbf{Exotic LSTs and T-polyality}: K3s with large Picard numbers
 generally admit  multiple inequivalent elliptic fibrations. This generalizes the order two T-duality of the $E_8\times E_8 \leftrightarrow Spin(32)/\mathbb{Z}_2$ heterotic strings to a polyality involving a number of 6D LSTs which equals the number of inequivalent elliptic structures on the given flavor K3 -- the same caveat in footnote \ref{WTF} applies here: the actual T-duality network of a given LST might be even larger if we allow for birational transformations of that type.
 \item \textbf{LSTs and generalized Kulikov-like degenerations:}
 Viewing the CY as a K3 fibration, the LST is localized at a locus where the K3 fiber degenerates in a way much similar to a Kulikov degeneration \cite{Kulikov_1977}. However, LSTs give a much finer classification than the original one, which is indicated by the possibility to realize all Kodaira types of fibers. We stress here that \textbf{the 2-group structure constants are invariants of these fibrations} and therefore give a different approach towards a classification scheme \textit{à la} Kulikov, but with labels given by the K3 frame lattice, $W$, the 5D Coulomb branch dimension of the LST, and the 2-group structure constant $\hat \kappa_{\mathscr R}$.
  \item \textbf{Flavor bounds and attractive K3s:}
  The frame lattice is a sublattice of $\Pi_{2,18}$ and it is bounded to have rank 18. The bound is saturated by the so-called attractive elliptic K3s. The models obtained from F-theory building on fibrations by attractive K3s generalize slightly the HW setup. Also in HW one obtains global symmetries at most of rank 18.\footnote{One has one $E_8$ factor from each M9 brane and an $SO(4) = SU(2) \times SU(2)$ from the directions transverse to the M5s.} For the models arising from attractive K3s one can obtain some exotic non-abelian enhancement of the flavor symmetry. When $G_F$ is maximal and non-Abelian, the underlying K3s are extremal. 
\end{enumerate}  
The structure of this review is as follows, in section \ref{sec:2groupreview} we give a short condensed review of the field theoretical results from generalized 6D quivers, as well as explaining explicitly how to compute 2-group structure constants. In section \ref{sec:FREW} we give a quick review of the F-theory geometries involved in these constructions. In section \ref{sec:DiscreteHolonomy} we give an explicit example of a T-duality and its geometric realization. In section \ref{ssec:ExtremalK3} we discuss the T-hexality arising from the different inequivalent elliptic structure on an extremal attractive K3.

\section{LSTs, 2-groups and Fusion}\label{sec:2groupreview}
In this section we recall the basic ingredients of how to engineer heterotic LSTs via non-compact elliptic threefolds. 
For a nice review see \cite{Heckman:2018jxk}
\medskip

First we denote the flavor lie algebra of a rank $n_T$  6D LST as
\be
\mathfrak f^{(0)} = \prod_{a=1}^{n_f} \mathfrak f_a\,,
\ee
where $\mathfrak f_a$ are irreducible factors. The dynamical degrees of freedom are denoted by a generalized quiver which is fixed by two sets of data  
\begin{itemize}
\item An $(n_T+1 + n_f) \times (n_T+1 + n_f)$ symmetric for the quiver 
\be
\left(\begin{matrix} \eta^{IJ} & \eta^{IA} \\ \eta^{AI} & 0\end{matrix}\right) \qquad \begin{aligned}& I,J = 1,...,n_T+1 \\ &A = 1,...,n_f \end{aligned}
\ee
 The block diagonals $\eta^{II}$ are positive integers that run from $0$ to $12$ and denote the Dirac pairings of the BPS strings that source the self-dual 2-form fields. All other entries are negative integers or vanishing.
\item A $(n_T+1 + n_f)$-tuple of Lie gauge algebras
\be
\mathfrak g = (\mathfrak g_1,..., \mathfrak g_{n_T+1}, \mathfrak f_1,..., \mathfrak f_{n_f})\,,
\ee
which result in dynamical couplings that couple to the $I$-th tensor fields, where $\eta^{II}$ is also identified with their anomaly coefficients.
\end{itemize}
Putting everything together the generalized quiver is given by assigning a node to each nonzero $\eta^{II}$ decorated with the Lie algebra $\mathfrak g_I$, and, whenever $\eta^{IA} \neq 0$, also with a factor $\mathfrak f_A$ of the symmetry Lie algebra. Schematically we write
\be
\cdots \, \underset{\atop [\mathfrak f_A]_{(\eta^{\text{\tiny{\textsc{ia}}}})}}{\overset{\mathfrak g_I \atop \,}{\eta^{II}}} \, \cdots
\ee 
The off-diagonal entries of the $\eta^{IJ}$ block are non-positive integers encoding the adjacency matrix for the generalized quiver: if $\eta^{IJ} \neq 0$ the two nodes $\eta^{II}$ and $\eta^{JJ}$ are neighbours and the corresponding BPS strings can form bound-states\footnote{
In this work the off-diagonal  $\eta^{IJ}$ are either $0$ or $-1$. More general cases are given in \cite{Bhardwaj:2015oru,Bhardwaj:2018jgp}. 
}. In terms of gauge and flavor algebra factors, such a non-trivial adjacency often yields bi-fundamental matter.

For 6D SCFTs the matrix $\eta^{IJ}$ must be positive definite, while for 6D LSTs it has to be positive semi-definite with a single unique vanishing eigenvalue. The LST is a linear combination of BPS bound states that precisely correspond to that vanishing eigenvalue. One might thus consider $\eta^{IJ}$ as a generalized \emph{affine} Cartan matrix) \cite{Bhardwaj:2015oru} such that
\be\label{eq:lstcharge}
\eta^{IJ} \ell_J = 0 \qquad \gcd(\ell_1,...,\ell_{n_T+1}) = 1 \qquad \ell_I > 0\,,
\ee
with $\ell_J$ is the generalized Kac label, identified as the LST $U(1)^{(1)}_{LS}$ charges of the BPS strings. 
 
Such quiver data is sufficient to fix the 6D theory and to obtain the 2-group structure constant given as
\be
  \framebox{$\phantom{  }\hat \kappa_\mathscr{F} =  - \sum_{I=1}^{r+1} \ell_I \eta^{IA} \qquad \hat \kappa_\mathscr{R} =   \sum_{I=1}^{r+1} \ell_I h^\vee_{\mathfrak{g}_I} \qquad \hat \kappa_\mathscr{P} = - \sum_{I=1}^{r+1} \ell_I (\eta^{II}-2) $}
   \ee
Those can be deduced by an analysis of the Bianchi identity of the non-dynamical LST 2-form background field \cite{Cordova:2018cvg,DelZotto:2020esg}, building on \cite{Ohmori:2014kda}.
In \cite{DelZotto:2020esg} it was first proposed and verified for many examples, that those structure constants are invariant across T-dual LSTs. In particular the matching of $\hat \kappa_\mathscr{P}$ and $\hat \kappa_\mathscr{R}$ yields strong novel conditions to match T-dual LSTs.

LSTs can be grouped into two classes reflected by the two possible values $0$ and $2$ of $\hat \kappa_\mathscr{P}$. For Heterotic LSTs $\hat \kappa_\mathscr{P}=2$. %The two types we call type II and heterotic LSTs respectively.
%In the following we want to focus on the later type leaving the former for future work.
   
As the name suggests, a good starting point to construct the second type of LSTs is the 10D $Spin(32)/\mathbb Z_2$ heterotic string.
In particular the worldvolume theories governing stacks of $N$ instantonic NS5 branes are well explored \cite{Witten:1995gx} and have a Lagrangian description with an $\mathfrak{sp}_N$ gauge algebra and $N_f = 16$ hypermultiplets in the fundamental and one in the antisymmetric representation. The corresponding generalized 6D quiver is
\be
[\mathfrak{so}(32)]  \overset{\mathfrak{sp}_N}{0}  [\mathfrak{su}(2)]\,.
\ee
which highlights the dual F-theory picture that we discuss momentarily.
Thanks to this Lagrangian description, the worldvolume theories of ALE heterotic $Spin(32)/\mathbb Z_2$ can be easily determined by orbifolding \cite{Blum:1997fw,Blum:1997mm,Intriligator:1997dh}. Since $\pi_1(S^3/\Gamma_{\mathfrak g}) \simeq \Gamma_{\mathfrak g}$, the resulting theories also depend on the choice of a flat connection at infinity, encoded in a choice of a mapping 
\be
\boldsymbol{\lambda}:\Gamma_{\mathfrak{g}} \to Spin(32)/\mathbb Z_2.
\ee
The singularity $\mathfrak{g}$ and a choice of flat connections\footnote{An important subtlety here is, that one could allow for boundary conditions without "vector-structure" in $Spin(32)/\mathbb Z_2$ via a positive second Stieffel-Whitney class $\widetilde{\textsf{w}}_2 \neq 0$ \cite{Berkooz:1996iz}. In this work we will only focus on configurations with vector structure leaving the other case for future work. 
} specifies the theory, which we denote by
\be
\widetilde{\mathcal K}_N(\boldsymbol{\lambda}; \mathfrak g) \, .
\ee 
The unbroken flavor $F(\boldsymbol{\lambda})$ group  of the 6D theory is determined by the commutant  of $\boldsymbol{\lambda}(\Gamma_{\mathfrak g})$ in $Spin(32)/\mathbb Z_2$. 

\medskip

Similarly one might consider the instantons on $E_8 \times E_8$. Such theories are closely related to the 6D $(2,0)$ SCFTs and hence have no conventional Lagrangian formulation. The Hořava-Witten formulation \cite{Horava:1995qa} yields a characterization of such theories: Compactifying M-theory on a finite interval with two 10D space-time filling M9 branes that carry an $E_8$ worldvolume theory. The LST arises from a stack of $N$ M5 branes that fill out six dimensions, parallel to the two M9 branes \cite{Ganor:1996mu}.  

The strings that stretch between M9-M5 brane are given by E-strings and those between M5-M5's are M-strings which we write \cite{Ganor:1996mu} as the generalized quiver  
\begin{align}
\label{eq:Instanton}
[\fe_8] \, \, 1 \, \, 2 \, \, 2 \ldots   2\, \, 2 \, \, 1 \, \, [\fe_8] \,. 
\end{align}
Along the residual four transverse directions to the M5 branes, an ALE singularity $\mathbb C^2/\Gamma_{\mathfrak g}$ is located. The presence of the singularity leads to a fractionalization \cite{Horava:1995qa,DelZotto:2014fia}, which also depends further on the choice of a flat connection at infinity for the two $E_8$ bundles. 
The latter are encoded in two group morphisms 
\be
\boldsymbol{\mu}_{a} \colon \pi_1(S^3/\Gamma_{\mathfrak g}) \simeq \Gamma_{\mathfrak g}\to E_8\,,
\ee
with $a=1,2$ labeling the two M9 branes. The zero form global symmetry of the resulting LSTs is determined by the commutant of $\boldsymbol{\mu}_a(\Gamma_{\mathfrak g})$ in $E_8$, namely we expect to have 
\be
F^{(0)}_a \equiv \{ g \in E_8 \, | \, gh = hg, \forall h \in \boldsymbol{\mu}_a(\Gamma_{\mathfrak g})\} \qquad a=1,2\,. \\
\ee
The cases with global symmetry $F_{1,2}^{(0)} \simeq E_8$ correspond to the choice of trivial flat connections for the end of the world gauge fields. They are dual to certain geometries in F-theory that had been discovered by Aspinwall and Morrison \cite{Aspinwall:1997ye}. In general, we denote the associated theories by
\be
\mathcal K_N(\boldsymbol{\mu}_1,\boldsymbol{\mu}_2;\mathfrak g)
\ee
The tensor branches of all these theories can easily be obtained by noting that each 1 and 2 curve piece is an SCFT in itself. The fractionalizations of those pieces have been determined in \cite{DelZotto:2014hpa} by exploiting F-theory techniques. 

%from the conformal matter picture by splitting up the chain \eqref{} into SCFT regions. I.e

 %The resulting fractions have been determined exploiting F-theory techniques  -- for an application in the context of this paper, we refer our readers to \cite{DZLO3}. % the above basis gets decorated as follows
%\be
%\underbrace{\overset{\mathfrak g}{1}\,\overset{\mathfrak g}{2}\,\overset{\mathfrak g}{2}\,\overset{\mathfrak g}{2}\, \cdots \, \overset{\mathfrak g}{2} \, \overset{\mathfrak g}{2} \, \overset{\mathfrak g}{1}}_{N+1}
%\ee
%but now the geometry needs several blow-ups to account for the collisions of discriminant loci supporting the $\mathfrak{g}-\mathfrak{g}$ gauge symmetries. 
The resulting theories are described as generalized quiver theories of the form
\be\label{eq:avheterotto}
\mathcal K_N(\boldsymbol{\mu}_1,\boldsymbol{\mu}_2;\mathfrak g) = \xymatrix{\mathcal T(\boldsymbol{\mu_1},\mathfrak g) \ar@{-}[r]^{\mathfrak {g}} &\mathcal T_{N-2}(\mathfrak g,\mathfrak g)  \ar@{-}[r]^{\mathfrak {g}}&\mathcal T(\boldsymbol{\mu_2},\mathfrak g) }
\ee
where 
\begin{itemize}
\item $\mathcal T(\boldsymbol{\mu_a},\mathfrak g)$ is the minimal 6D orbi-instanton theory associated to the M9-M5 brane system system in presence of a $\mathbb C^2/\Gamma_{\mathfrak g}$ singularity transverse to the M5, includng a choice of flat connection at infinity $\boldsymbol{\mu_a}:\Gamma_{\mathfrak g} \to E_8$.
\item $\mathcal T_{N-2}(\mathfrak g,\mathfrak g)$ is the 6D conformal matter theory associated to $N-2$ M5 branes probing a $\mathbb C^2/\Gamma_{\mathfrak g}$ singularity;
\item $\xymatrix{\ar@{-}[r]^{\mathfrak g}&}$ denotes the operation of (diagonal) fusion of the common factors $\mathfrak g$ of the global symmetry of the corresponding 6D SCFTs, schematically at the level of the corresponding generalized quivers:
%\footnote{This is the 6D version of the gauging operation in 4d, which our readers are probably more familiar with. For further references about this, see \cite{DelZotto:2018tcj} and \cite{Heckman:2018pqx}. Our readers that are not familiar with this operation can find plenty of examples in the discussion below.}
\be
\xymatrix{\cdots \,\, \overset{\mathfrak g'}{n'} \,\, \textcolor{red}{[\mathfrak g]} \ar@[red]@{-}[r]^{\textcolor{red}{\mathfrak g}}\,\,& \textcolor{red}{[\mathfrak g]}   \,\,  \overset{\mathfrak g''}{n''} \cdots }\,\, \longrightarrow \,\,\cdots \overset{\mathfrak g'}{n'} \,\,\textcolor{red}{\overset{\mathfrak{g}}{n}} \,\, \overset{\mathfrak g''}{n''} \cdots
\ee
\end{itemize}
The theories $\mathcal T(\boldsymbol{\mu_a},\mathfrak g)$ 
have been classified in the literature  \cite{DelZotto:2014hpa,Heckman:2015bfa,Mekareeya:2017jgc,Frey:2018vpw}, by identifying the minimal orbi-instanton model. When $N \geq 3$, the structure in \eqref{eq:avheterotto} completely determines the tensor branches of all possible fractional $E_8 \times E_8$ heterotic instantons for all possible singularities and holonomoies $\mu_a$, complementing the results available for this class of models in the literature \cite{Aspinwall:1996vc,Aspinwall:1997ye,Intriligator:1997dh}.
In \cite{DelZotto:2022ohj} we explain a strategy to match all $Spin(32)/\mathbb{Z}_2$ instantonic LSTs with holonomy $\boldsymbol{\lambda}$ in terms of the $E_8 \times E_8$ duals with holonomies $\boldsymbol{\mu}_a$ for all singularities $\mathfrak{g}$. We focus on the cases with exceptional flavor symmetries and gauge algebras, which cannot be realised in perturbative strings. For those case we give a complete characterization of the simply laced cases.

 \section{Heterotic LSTs as elliptic K3 Degenerations}\label{sec:FREW}
 In this section we want to use the geometric engineering framework of F-theory, to describe heterotic LSTs via   elliptic CY threefolds 
  \begin{align}
\begin{array}{cc}
T^2 \rightarrow &X_3 \\
&  \downarrow \pi\\
& B_2
\end{array}
 \end{align}
over non-compact bases $B_2$. As stated in the introduction this requires to have a base $B_2$ that shrinks to a curve $\Sigma_{LS}$. Such curve typically is not irreducible. For Heterotic LSTs the irreducible components of $\Sigma_{LS}$ are always rational curves intersecting transversally. Then $\eta^{IJ}$ is identified with the intersection pairing $\eta^{I J} = -w^I \cdot w^J$ of the base $B_2$ using
$w^I \in H_2(B_2, \mathbb{Z})$ to denote the irreducible components of $\Sigma_{LS}$. The gauge algebra factors $\mathfrak g_I$ correspond to Kodaira singularities of the elliptic fiber along the irreducible component $w^I$, the irreducible components of the flavor symmetry algebra are encoded in the Kodaira types of the non-compact components of the discriminant locus of the elliptic fibration.  It is then easy to show, that the 2-group structure constant $\hat \kappa_\mathscr{P}$ is a birational equivalent of the base $B_2$ and in particular $\hat \kappa_P=2$ leads to base of the schematic form
 \begin{align}
 \label{eq:BaseHet}
B_2 =  BL^{n_T}( \mathbb{P}^1 \times \mathbb{C} ) \, ,
 \end{align}
 where we stress that most blow ups occur at non-generic points.
The LST  curve is given as the linear combination dictated by \eqref{eq:lstcharge}:
\begin{align}
\Sigma_{LS} = \sum_{I=1}^{n_T+1} \ell^I  w^I \, ,
\end{align}
which is by construction a rational curve of null self-intersection $\Sigma_{LS}^2=0$. This directly implies that for Heterotic LSTs the threefold $X_3$ also admits the structure of an elliptic K3 fibration
\begin{align}
\begin{array}{cc}
K3 \hookrightarrow& X_3 \\
& \downarrow \pi \\
& \mathbb{C}_u  
\end{array} \, , 
\quad \text{ with } \quad
\begin{array}{cc}
T^2 \hookrightarrow& K3 \\
& \downarrow \pi \\
& \Sigma_{LS} 
\end{array} \, .
\end{align}
This is directly clear for $n_T=0$ where $\Sigma_{LS}=\mathbb{P}^1$ in \eqref{eq:BaseHet}. We proceed by explaining first the properties of the K3 fiber encoding the flavor symmetry and then the compact divisors at the origin.  

\textbf{The Flavor K3:} We denote the $K3$ fiber by $S$. 
By embedding the K3 into the whole threefold we must fix a polarization of $S$ which fixes the N$\acute{\text{e}}$ron-Severi lattice NS$(S)$ as a sublattice of $\Lambda_{K3}=II_{3,19}$
given as
\begin{align}
 NS(S) := H^{1,1}(S) \cap H^2(K3, \mathbb{Z}) \, .
 \end{align}
The requirement for the K3 to be elliptic fixes more structures of NS$(S)$ namely to write it as
\begin{align}
NS(S) = U \oplus W \, ,
\end{align}
where $U$ is the hyperbolic lattice corresponding to the section and the base, $W$ is the frame lattice. The frame lattice is spanned by divisors that do not intersect the zero-section. 
The NS$(S)$ lattice connects the K3 with the non-compact divisors $D \in X_3$ that we identify with the resolution divisors of the flavor group.

In the following we assume that $X_3$ admits additional inequivalent fibration structures 
\begin{align}
\begin{array}{cc}
T^2 \rightarrow &  X_3 \\
 & \downarrow \tilde{\pi} \\ 
 & \hat{B}_2
 \end{array}
\end{align}
with a new base $\hat{B}_2$ and potentially non-trivial fibers. The projection to the new base $\hat{B}_2$ however, must hold for a generic point of the non-compact base $\mathbb{C}_u$ far away from the compact divisors. This locus however, is spanned by the non-compact divisors $D \in NS(S)$ which pull back to the K3 $S$ under $\tilde{\pi}$ as well. This implies that all inequivalent fibrations of $X_3$ must respect the K3 fibration. Hence all T-duals with a geometric origin from inequivalent elliptic structures of $X_3$ must arise from inequivalent elliptic structures of the K3 fiber. Hence these are automatically heterotic LSTs for which we argued that $\hat \kappa_\mathscr{P}=2$. We have thus proven geometrically the invariance of $\hat \kappa_\mathscr{P}$ under T-duality. Moreover we have reduced the problem of finding all inequivalent elliptic fibrations of the non-compact threefold $X_3$ to those of the compact K3 $S$ with fixed polarization. Finding all elliptic fibrations for a polarized K3 is still a hard problem but the self-dual property of the K3 lattice structure mitigates this problem substantially.  
\\\\
 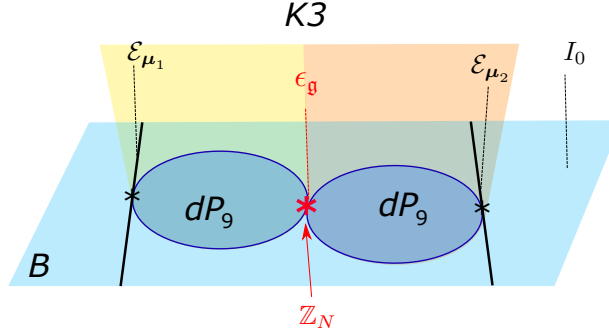
\begin{figure}[ht!]
	 \begin{center}
  \begin{picture}(0,140)
	 \put(-110,30){\includegraphics[scale=0.5]{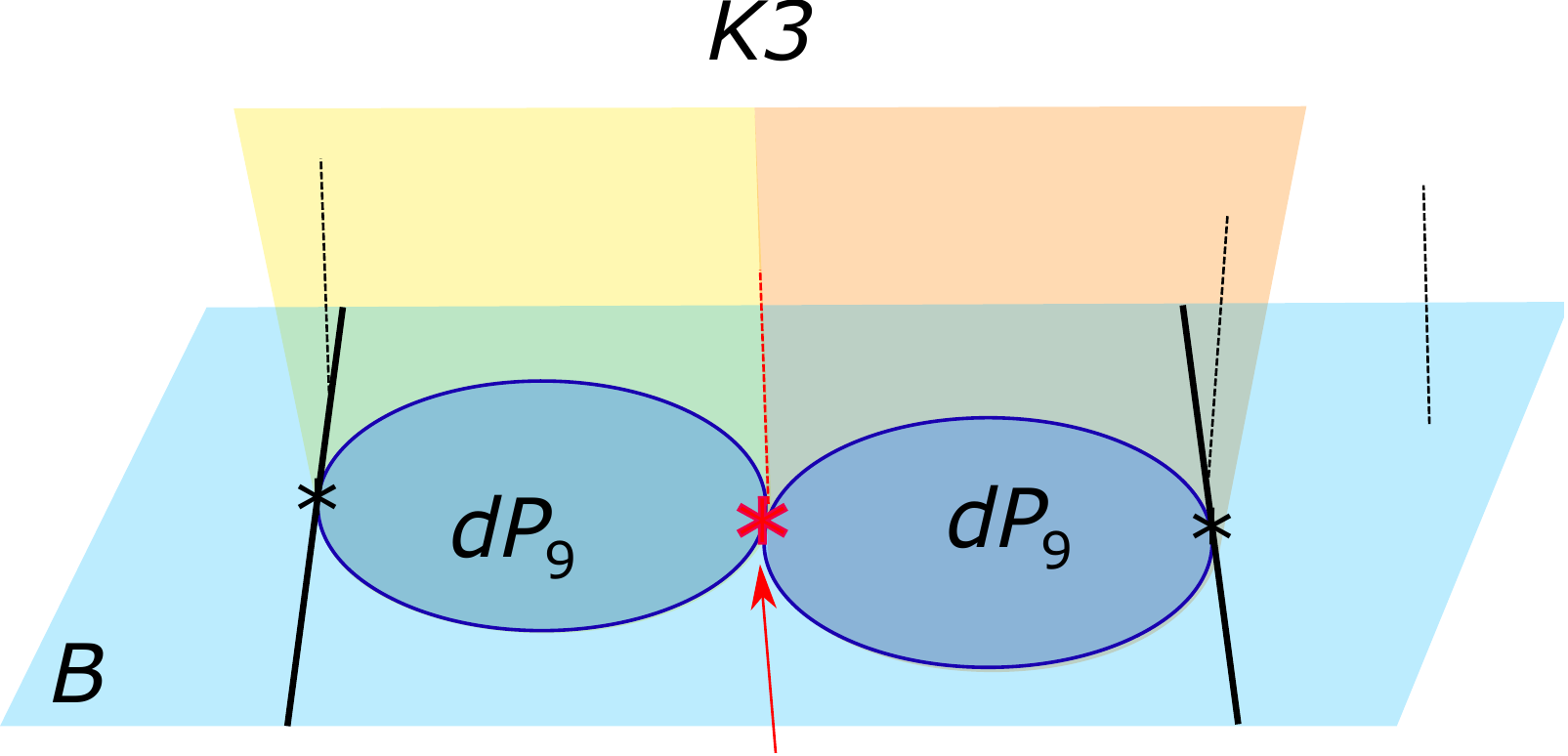}}
   \put(0,20){\textcolor{red}{$\mathbb{Z}_N$}}
    \put(-65,120){$\mathcal E_{\boldsymbol{\mu}_1}$}
    \put(65,115){$\mathcal E_{\boldsymbol{\mu}_2}$}
    \put(100,120){$I_0$}
     \put(-5,110){ {\Large \textcolor{red}{$ \epsilon_\mathfrak{g} $}}}
  \end{picture}
	 \caption{ 
Heterotic LST from a stable degeneration of the K3 into two rational elliptic surfaces. At the intersection locus of the two $-1$ curves in the base, is a $\mathbb{Z}_{N=n_T-1}$ singularity with another Kodaira type $\mathfrak{g}$ singularity on top. Flavor holonomies $\boldsymbol{\mu}_i$ for the  two  
Hořava-Witten walls, dualize to  $\mathcal E_{\boldsymbol{\mu}_a}$ Kodaira fibers of the rational elliptic surface. }\label{fig:K3deg}
	 \end{center}
	  \end{figure}
\textbf{The degeneration Locus}
The compact divisors are located at the locus $u=0$ of the base $\mathbb{C}$. Note that for $n_T=1$ with no singularities  the base  quiver takes the form
 \begin{align}
 1 \, \, 1
 \end{align}
 This can be interpreted geometrically as follows. The restriction of the $T^2$ fibration along the -1 base curves gives a rational elliptic surfaces (the so-called del Pezzo 9) into which the K3 fiber has degenerated into, see Figure \ref{fig:K3deg}. More in general note that the quiver given in \eqref{eq:Instanton} is nothing but a Kulikov degeneration of type II\footnote{See \cite{Kulikov_1977} for a formal definition}. Note that we can shrink the -2 curves to yield a $\mathbb C^2 / \mathbb{Z}_{n_T-1}$ singularity.

In a similar way, the two non-trivial ingredients, discussed in Section~\ref{sec:2groupreview} admit a geometric counterpart: 

  First decorating the quiver with gauge groups $\mathfrak{g}$ is achieved 
by taking a singularity $u=0$ transversely, which yields the same type of singular elliptic fibers. Secondly, non-trivial flavor holonomies are obtained by rotating the flavor K3 to a smaller NS lattice.  
 
\begin{figure}[t!]  
  \begin{center}
	 \begin{picture}(0,200)
\put(-142,20){\includegraphics[scale=0.9]{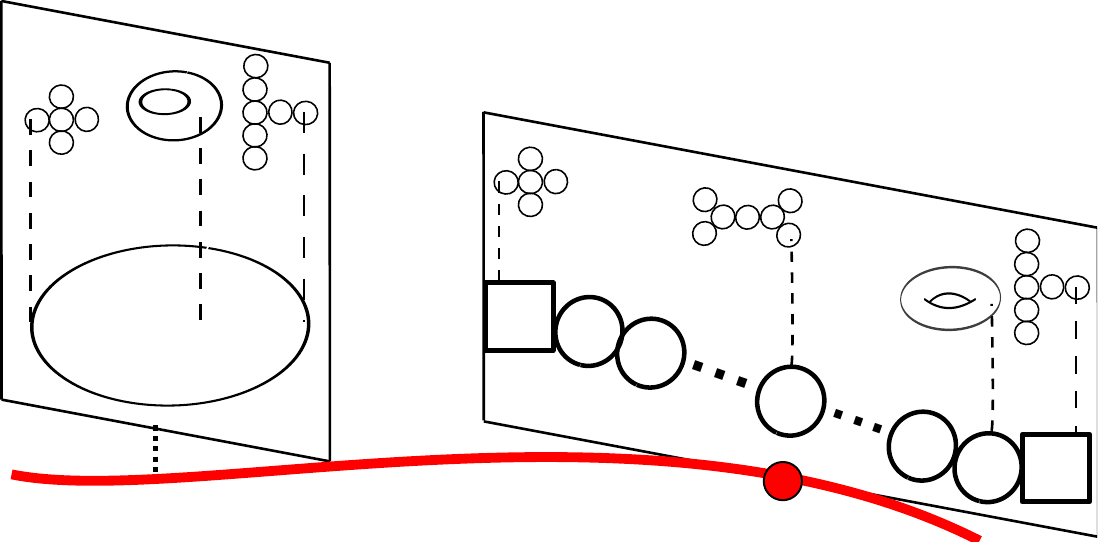}}
\put(-70,150){K3}
\put(125,40){$\mathfrak{g}_{F_2}$}
%\put(70,80){$\mathfrak{g}_{F_1}$}
\put(-20,150){"generalized" Kulikov -Degeneration}
\put(-30,25){\LARGE $\rightarrow$}
\put(-90,20){$\textcolor{red}{\mathbb{C}_u}$}
\put(50,20){$\textcolor{red}{u=0}$}
\put(-78,75){$\mathfrak{g}_{F_2}  $}
\put(-14,80){$\mathfrak{g}_{F_1}  $}
\put(-130,80){$\mathfrak{g}_{F_1}  $}
\end{picture}
	 \caption{A little string theory as a "Kulikov-like" degeneration: Over a generic point in the base $\mathbb{C}_u$ there is an elliptic K3 fiber, whose frame lattice yields the flavor group. At the origin of $\mathbb{C}$, the K3 degenerates into compact surfaces with $\mathfrak{g}$ fibers corresponding to a fusion of $\mathcal{T}(\mathfrak{g},\mathfrak{g}_{F_i})$ conformal matter.}\label{fig:Kulikov}
 \end{center}
 \end{figure}
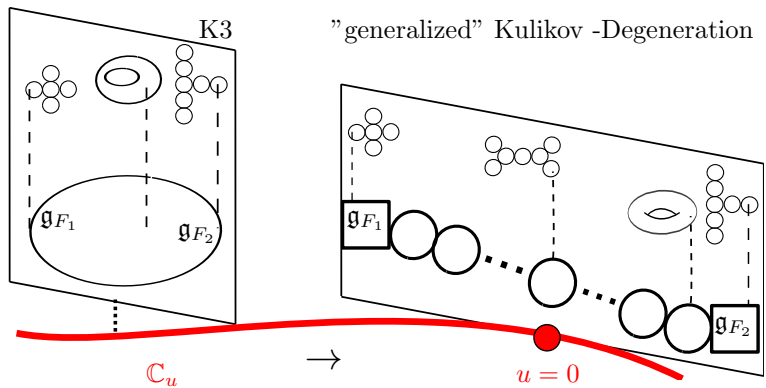

Such type of singularities and their resolutions, go beyond Kulikov degenerations. The structure however is similar enough to motivate a sufficient generalization of K3 degenerations that captures the full set of Heterotic LSTs mathematically. The full K3 degeneration picture of the threefold that corresponds to Heterotic LSTs is schematically depicted in Figure~\ref{fig:Kulikov}. 

The fact that Heterotic LST dualities must descend from inequivalent elliptic fibration structures of the flavor K3, leads to a natural algorithm to find all duals for a Heterotic LST arising from a single chamber of the 5D Coulomb moduli space: 
\begin{enumerate}
    \item  Construct an arbitrary heterotic LST and determine its flavor group $G_F = \prod_A \mathscr F_A$;
    \item Find the corresponding flavor K3 that realizes $G_F$ as its frame lattice.
    Determine all inequivalent fibrations using lattice techniques and determine their frame lattices. This yields the collection of all possible dual flavor groups $\{G^(e)_F\}_{e \in E(S)}$ compatible with the given $X_3$, ie. realised within the same chamber of the 5D moduli space. Here $e$ is an index running over the possible elliptic structures $\mathbb E(S)$ of $S$;
    \item Starting from the SCFT classification results, construct a family of dual LSTs labeled by elements in $\mathbb E(s)$, obtained by fusing SCFTs with the constraints that \textit{(i)} the dual flavor groups $G^{(e)}_F$ are compatible,\footnote{ This requires not only that the corresponding ranks are the same, but also that they have matching full subalgebras, and compatible $\hat \varkappa_{\mathscr{F}_A}$ -- sometimes this requires rescaling those by the index of embedding.} \textit{(ii)} the 2-group structure constants match, and \textit{(iii)} the 5D Coulomb branch dimension are the same.
\end{enumerate}
We discuss an explicit example of this strategy at work to construct six dual LSTs in Section~\ref{ssec:ExtremalK3}.

\section{Example 1: Discrete Holonomy LSTs}\label{sec:DiscreteHolonomy}
In this section we exemplify the above considerations, using a modified version of Batyrev's construction to obtain the resolved non-compact threefold as a hypersurface in a non-compact ambient fourfold.

When doing so, we will start with a theory on the $E_8 \times E_8$ side, adding a singularity and flavor holonomies at infinity. Upon the toric resolution, we will read off not only the T-dual on the $Spin(32)/\mathbb{Z}_2$ side, but also a third exotic dual Heterotic LST.

Starting with the $E_8^2$ we take $N+1$ NS5 branes with an $E_7$ singularity.
The $\mathbb{Z}_2$ center of $E_7$ allows to choose a discrete holonomy $\mu_a=\mathbb{Z}_2$  which yields a breaking to the unbroken flavor group
\begin{align}
G_F:  E_8^2 \rightarrow  (E_7 \times SU(2))^2/ \mathbb{Z}_2  \, .
\end{align}
Note that this choice yields a non-simply connected flavor group 
\cite{Dierigl:2020myk,Aspinwall:1998xj}.
Upon fractionalization, the full quiver then becomes  
\begin{align} 
\label{eq:e7Quiver}
\begin{array}{c } 
\, \, {\overset{ \lbrack\mathfrak{su}_2\rbrack}{1}}\, \,   \qquad \qquad \qquad \qquad \qquad \quad \qquad \qquad {\overset{ \lbrack\mathfrak{su}_2\rbrack}{1}} \, \,  \\ 
\lbrack \fe_7\rbrack  \,\,  1 \,  {\overset{\mathfrak{su}_2}{2}}\,\, {\overset{\mathfrak{so}_7}{3}} \,\, {\overset{\mathfrak{su}_2}{2}} \, \, 1\, \,  \underbrace{\overset{\mathfrak{e}_7}{8} \,\,  1 \,  {\overset{\mathfrak{su}_2}{2}}\,\, {\overset{\mathfrak{so}_7}{3}} \, \,  {\overset{\mathfrak{su}_2}{2}} \, \, 1 \, \,   \overset{\mathfrak{e}_7}{8}  \ldots \overset{\mathfrak{e}_7}{8}  \,\,  1 \,  {\overset{\mathfrak{su}_2}{2}}\,\, {\overset{\mathfrak{so}_7}{3}} \,\, {\overset{\mathfrak{su}_2}{2}} \, \, 1\, \,  \overset{\mathfrak{e}_7}{8}}_{\times N} \,\,  1 \,  {\overset{\mathfrak{su}_2}{2}}\,\, {\overset{\mathfrak{so}_7}{3}}  {\overset{\mathfrak{su}_2}{2}} \, \, 1 \, \,   
 \lbrack \fe_7\rbrack    \, .
\end{array}
\end{align}
The repeating chain  $ 
1 \,  {\overset{\mathfrak{su}_2}{2}}\,\, {\overset{\mathfrak{so}_7}{3}} \, \,  {\overset{\mathfrak{su}_2}{2}} \, \, 1 \, \,  $ 
is the $\mathcal{T}(\mathfrak{e}_7,\mathfrak{e}_7)$ conformal matter required from the fractionalization of the $M5$ branes. This data is sufficient to compute the dimension of Coulomb branch. Note that each of the $N$ chains contributes 6 tensors and an $\fe_7 \times \fsu_2^2 \times \fso_7$ gauge group and hence contributes with $18$ Coulomb branch parameters. From this we deduce the general expression
\begin{align}
\text{dim(CB)}= \mathbf{T}+\text{rk}(G)= 11+18N \, .
\end{align} 
Next we require the LS charge vector in order to compute the 2-group structure constants
\begin{align}
\begin{array}{c}
\qquad \, \, 1,   \,\, \quad \qquad \qquad \qquad \quad \qquad \qquad 1,\\
\vec{l}_{LS}=(1, 1, 1, 2, 3, \underbrace{1, 4, 3, 2, 3, 4, 1 \ldots 1, 4, 3, 2, 3, 4, 1}_{\times N}, 3, 2, 1, 1, 1  )   \, ,
\end{array}
\end{align}
which we depict in the same form as the quiver to simplify identification of LS charges with the quiver nodes.

In order to compute   $\hat \kappa_\mathscr{R}$ we also require the dual 
Coxeter numbers $h^\vee_{\mathfrak{g}}$ of the gauge algebra factors $\mathfrak{g}_I$ in the quiver. We will also write them in terms of a vector as 
\begin{align}
\begin{array}{c}
\qquad   1,   \quad \qquad \qquad \qquad \qquad \quad \qquad \qquad 1, \, \,\\
\vec{h}^\vee_{\mathfrak{g}_I}=(1, 2, 5, 2, 1, \underbrace{18, 1, 2, 5, 2, 1, 18 \ldots 18, 1, 2, 5, 2, 1, 18}_{\times N}, 1, 2, 5, 2, 1,  )   \, ,
\end{array}
\end{align}
Note that we assign a non-trivial dual Coxeter number to trivial gauge groups  as $h^\vee_{\fsu_0}=1$. From the repeating chain, we can then compute the 2-group contribution for each $\fe_7$ and conformal matter as
\begin{align}
\hat \kappa_\mathscr{R}(\overset{\fe_7}{8} \oplus \mathcal{T}(\mathfrak{e}_7,\mathfrak{e}_7)) =  (1,4,3,2,3,4)^T \cdot (18, 1, 2, 5, 2, 1)=48 \, .
\end{align}
Including $N$ copies of the above contribution to $\hat \kappa_\mathscr{R}$ we obtain for the full quiver 
\begin{align}
    \hat \kappa_\mathscr{R} = 2+48N \, .
\end{align}
We summarize the full data of the theory: This includes the frame lattice $W$ of the K3, which encodes the flavor group, the Coulomb branch dimension and 2-group structure constants as
\begin{align}
   (W, \, \text{dim(CB)},\, \hat \kappa_\mathscr{R} , \, \hat \kappa_\mathscr{P} ) \, =  \,  (\frac{E_7^2 \times SU(2)^2 }{\mathbb{Z}_2},\, 11+18N,\, 2+48N, \, 2 ) \, . 
\end{align} 

To deduce the T-duals, we explicitly resolve the above family of elliptic threefolds, starting with the compact flavor K3.
The global structure of the flavor group is implemented by a $\mathbb{Z}_2$ Mordell-Weil (MW) group.  

The K3 is specified by a 3d reflexive lattice polytope, with points given in Table~\ref{tab:K3Poly}. The elliptic K3 hypsersurface can easily be constructed using the Batyrev prescription, (see \cite{Batyrev:1994pg} for more details) which generalizes to the non-compact as well.
\begin{center}
\begin{table}[t!]
\begin{tabular}{cc}
\begin{tabular}{|c|c| }\hline  
\multicolumn{2}{|c|}{fiber 1} \\ \hline
Z & (1,0,0)     \\
X & (-1,2,0) \\
Y& (-1,-2,0) \\ 
$e_1$& (0,-1,0)\\  \hline
\end{tabular} 
&
\begin{tabular}{|c|c| }\hline  
\multicolumn{2}{|c|}{$E_7$ tops} \\ \hline
$\beta_{1,\pm} $ & $(-1,2,\pm 1  )$ \\
$\beta_{2,\pm} $ &$ (-1, 2,\pm 2  )$ \\ 
$\beta_{3,\pm} $ & $(-1,2,\pm 3  )$  \\ 
$\beta_{4,\pm} $ & $(-1,2,\pm 4  )$  \\
$\alpha_{3,\pm} $ &$ (-1,1,\pm 3  )$ \\    
$\alpha_{2,\pm} $ & $(0,1,\pm 2  )$  \\ 
$\alpha_{1,\pm} $ & $(-1,1,\pm 1  )$  \\
$\gamma_{2,\pm} $ &$ (-1,0,\pm 2 )$  \\ \hline   
\end{tabular}
\end{tabular}
\caption{\label{tab:K3Poly}Depiction of the toric rays of the flavor K3 polytope. 
The polytope admits three, 2d sub-polytopes which yield inequivalent elliptic fibrations with $X,Y,Z,e_1$ spanning the first. We have also highlighted the rays of the two $\fe_7$ tops.  
}
\end{table}
\end{center}
The elliptic fibration of the K3 is evident from its 2d sub-polytope structure spanned by $X,Z,Y,e_1$. This sub-polytope highlights a property of a the ambient variety which itself admits a fibration by a 2d complex Fano variety. The torus fibration is thus inherrited from the ambient space variety. \footnote{See \cite{Anderson:2016cdu,Huang:2019pne} for more details on this construction for compact threefolds.}

The residual rays resolve two $E_7$ singularities \cite{Candelas:1996su,Bouchard:2003bu}.
Note that this fiber structure, admits a $\mathbb{Z}_2$ MW group which implements the non-trivial global structure \cite{Aspinwall:1998xj,Mayrhofer:2014opa,Klevers:2014bqa}.

This is the first building block to construct a non-compact threefold. For this we consider a 4d polytope $\Delta_4$, in which the K3 is given as a 3d  reflexive sub-polytope $\Delta_3$  as
\begin{align}
\Delta_4= \left(\begin{array}{c|c} 
\Delta_3 & 0\\ \hline
\multicolumn{2}{c}{
\hat{\Delta}_4}
\end{array}
\right)\, .
\end{align}
The compact surfaces are specified be the collection of  vertices given by the collection $\hat{\Delta}_4$, spelled out in detail in Table~\ref{tab:CompactE7} in the Appendix and is sufficient to deduce the resolved quiver in \eqref{eq:e7Quiver}.
 
By deducing two more 2d sub-polytopes in the 3d sub-polytope of the K3 we deduce the other toric fibrations, their flavor tops and intersections structure of the compact divisors. 
The second 2d sub-polytope is spanned by the vertices $Z,\gamma_{2,+}$ and $\gamma_{2,-}$ which also admits a $\mathbb{Z}_2$ Mordell-Weil group. 

After re-arranging all non-compact and compact divisors, we obtain the quiver 
\begin{align}
\begin{array}{c} 
  {\overset{\mathfrak{sp}_{2N-4}}{1 }} \, \, \, \, \, \,  \\ 
\lbrack \fso_8\rbrack    \, \,  {\overset{\mathfrak{sp}_{N-1}}{1}} \, \,  {\overset{\mathfrak{so}_{4N+4}}{4}} \, \, {\overset{\mathfrak{sp}_{3N-3}}{1}} \, \,   {\overset{\mathfrak{so}_{8N }}{4 }} \, \,    {\overset{\mathfrak{sp}_{3N-1}}{1}} \, \,    {\overset{\mathfrak{so}_{4N+12}}{4}} \, \,   {\overset{\mathfrak{sp}_{N+5}}{1}} \, \,   \lbrack \fso_{24}\rbrack  \, ,
\end{array}
\end{align}
with LS charges 
\begin{align}
\begin{array}{c}
\qquad \, \, \,  2, \\
\vec{l}_{LS}=(1,1,3,2 ,3,1,1 )   \, .   
\end{array}
\end{align}
Note that we find this dual LST quiver to admit the topology of an $\fe_7^{(1)}$, which is just the singularity that appeared in the first fibration. Computation of the Coulomb branch is simpler here as the number of tensors is fixed to $T=7$ which yields
\begin{align}
    \text{Dim(CB)}= T + \text{rk} (G) =11+18N \, .
\end{align}
Using for the dual Coxeter numbers
\begin{align}
h^\vee_{\fso_{2M}}=2M-2 \quad \text{ and }\quad h^\vee_{\fsp_M}=M+1 \, ,
\end{align}
we compute the 2-group structure constants that are summarized in the full data of the theory  
\begin{align}
   (W,\,  \text{dim(CB)},\,  \hat \kappa_\mathscr{R} ,\, \hat \kappa_\mathscr{P} ) \, =  \,  (\frac{\text{Spin}(24)\times \text{Spin}(8)}{\mathbb{Z}_2},\, 11+18N,\, 2+48N, \, 2 ) \, . 
\end{align}

The third 2d sub-polytope is spanned by $e_1, \alpha_{2,+}$ and $\alpha_{2,-}$. This model admits also a $\mathbb{Z}_2$ group in addition to a free part which yields a $\mathfrak{u}_1$ flavor contribution \cite{Lee:2018ihr,Apruzzi:2020eqi}. Re-arranging all curves then yields the following quiver  
\begin{align}   
\lbrack \mathfrak{u}_{16} \rbrack    \, \,  {\overset{\mathfrak{su}_{2N+10}}{2}} \, \,  {\overset{\mathfrak{su}_{4N+4}}{2}} \, \, {\overset{\mathfrak{su}_{6N-2}}{2}} \, \,   {\overset{\mathfrak{sp}_{4N-4}}{1}} \, \,   {\overset{\mathfrak{so}_{4N+4}}{4 }} \, ,
\end{align}
for which we have the LS charge vector  
\begin{align}
	\vec{l}_{LS} = (1,2,3,4,1 ) \, .
\end{align} Note that the shape of this quiver does not resemble the shape of an affine $ADE$ quiver but is closer to an $\fe_6^{(2)}$ (but with symmetric intersections.). 
Using $h^\vee_{\fsu_M}=M$, Coulomb branch and 2-group structure constants can be readily computed.

The full data of the theory is summarized as 
\begin{align}
   (W, \,  \text{dim(CB)}, \,  \hat \kappa_\mathscr{R} , \,  \hat \kappa_\mathscr{P} ) \, =  \,  ( 
   \frac{SU(16) \times U(1)}
   {\mathbb{Z}_2},\, 11+18N,2+48N,\, 2 ) \, . 
\end{align}
 
We have proven the explicit duality of all three LSTs. As expected, we can match in all three models an explicit match of flavor rank, Coulomb branch dimension, and 2-group structure constants that we summarize as 
\begin{align}
(\text{rk(G$_F$)}, \, 
\text{Dim}(\text{CB}), \, 
\widehat{\kappa}_{\mathscr R}, \,
\widehat{\kappa}_{\mathscr P} )=(16, \,  18N+11 , \,  48N+2,\,  2) \,.
\end{align} 

 \section{Example 2: Extremal Flavor Rank and T-Hexality}
\label{ssec:ExtremalK3}

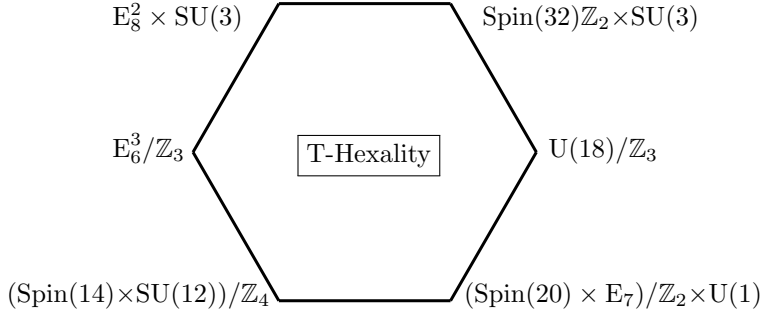
\begin{figure}
\begin{center}
    \begin{picture}(0,130)
    \put(-75,10){\includegraphics[scale=0.4 ]{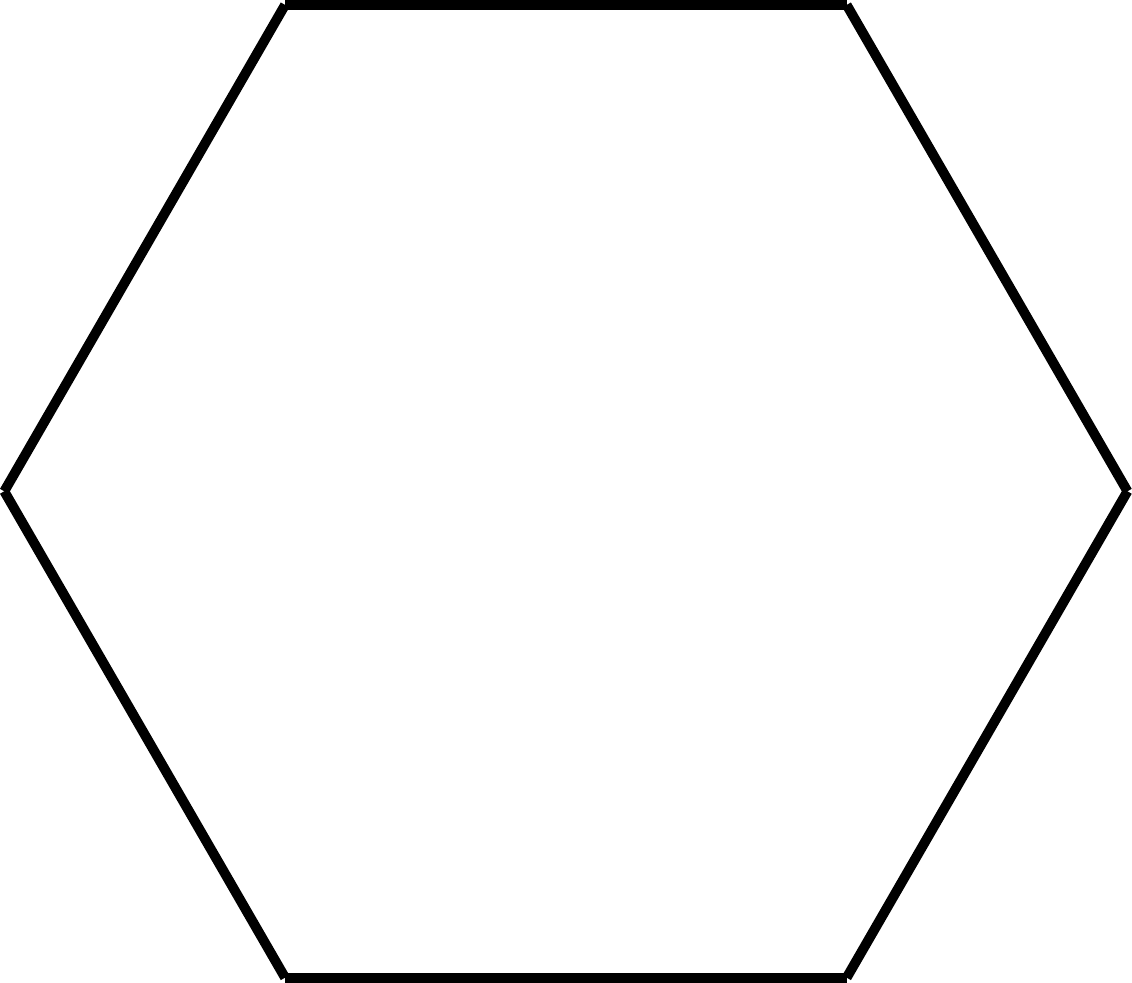}}
    \put(-105,115){E$_8^2 \times \text{SU}(3)$}
    \put(35,115){$\text{Spin}(32) \mathbb{Z}_2 \times $SU$(3)$}
    \put(-105,65){$\text{E}_6^3/\mathbb{Z}_3$}
     \put(60,65){U$(18)/\mathbb{Z}_3$}
     \put(-145,10){$(\text{Spin}(14)\times $SU$(12))/\mathbb{Z}_4$}
     \put(28,10){$(\text{Spin}(20)\times \text{E}_7)/\mathbb{Z}_2\times$U$(1)$}
     \put(-35,63){\framebox{T-Hexality}}
    \end{picture}
\end{center}
\caption{\label{fig:K3Hexagon}
The six different flavor groups, obtained from inequivalent elliptic fibration structures of the same attractive K3 \cite{Braun:2013yya}. Those map to non-geometric heterotic string backgrounds and yield the universal flavor groups of six different 6D LSTs.
}
\end{figure}

Using toric geometry in the section before, we showed that we 
often obtain more than two dual heterotic LSTs. This might come as a surprise, as one might think of this duality to descend from the familiar 10d $E_8\times E_8 \leftrightarrow Spin(32)/\mathbb{Z}_2 $. However, there might be even more fibrations that we could not access in the toric construction employed in the section before. In this section we want to go beyond those restrictions and propose a T-hexality of LSTs that admit the maximal flavor rank 18. 
%Such large flavor groups on the other hand, might also come as a surprise when one thinks of the LSTs in terms of the HW picture and expects it to be of rank 17 at most. 

From the geometry these large flavor groups descend from the frame lattice of the K3
which can be as large as rank 18.\footnote{ This suggests, that the actual duality of the LSTs originates from the eight dimensional SUGRA, where F-theory on K3 is dual to the heterotic string on $T^2$ and the flavor symmetry is a gauge symmetry. It would be interesting to reinterpret the 6d Heterotic LSTs as codimension 2 membranes for the 8d SUGRA - the Hagedorn density of states of LSTs might have some interesting implications for the physics of those systems.} There are plenty of examples in the mathematics literature:
Elliptic K3s with a maximal non-Abelian group of rank 18 are called extremal and have been classified in \cite{Miranda1986,2005math......5140S}. None of them have a toric construction, which makes a full resolution more complicated. The rigid structure of those K3s on the other hand makes a full classification of their elliptic fibrations in lattice theoretic terms feasible \cite{Braun:2013yya}.

In the following we exploit this fact and make use of the geometric algorithm outlined in Section~\ref{sec:DiscreteHolonomy} using an attractive K3 where all six inequivalent fibrations are known, see Figure~\ref{fig:K3Hexagon} which we use to propose dual LSTs.
For all six theories, rank of Flavor group, Coulomb branch dimension and 2-group structure constants 
\begin{align}
(\text{rk(G$_F$)}, \, \text{Dim(CB)}, \,   \widehat{\kappa}_{\mathscr R} , \, \widehat{\kappa}_{\mathscr P}) \, =\,  (18, 21, 30, 2) \, , 
 \end{align} 
match which we take as strong evidence for the existence of the duality. In what follows we list the 6 LSTs we identified.

 \paragraph{{\bf LST 1. $\mathbf{E_8^2 \times SU(3)}$ Flavor: }}
 This configuration is very close to the HW setup, but admits an additional $SU(3)$ factor. This factor can be deduced by employing the 8D SUGRA theory, corresponding to the heterotic string compactified on a self-dual $\fsu_3$ torus.
Using this 8D gauge group as a flavor group and conformal matter techniques \cite{DelZotto:2014fia} we deduce the following consistent LST quiver  
\begin{align}  
\begin{array}{ccc}
 \lbrack \fsu_3\rbrack  \\
 1 \\ 
\lbrack \fe_8\rbrack \, \, 1   \, \,2 \, \,   {\overset{\mathfrak{sp}_1}{2}}  \, \,   {\overset{\mathfrak{g}_2}{3}} \, \, 1 \, \,   \overset{\ff_4}{5}   \, \, 1 \, \,    {\overset{\mathfrak{g}_2}{3}}  \, \,   {\overset{\mathfrak{sp}_1}{2}}  \, \,  2 \,  \,  1 \,  \lbrack \fe_8\rbrack 
\end{array} \, ,  \qquad 
\begin{array}{c }
 \quad \, \, \, 1, \\ 
 \vec{l}_{LS}= (  1,1,1,1,2,1,2,1,1,1,1,1) \,.
\end{array} \, 
\end{align} 
\paragraph{{\bf LST 2. $\mathbf{ SO(32)/\mathbb{Z}_2 \times SU(3)}$ Flavor:}}When compared to the model before, this 
 case reproduces the familiar 10D heterotic duality upon which the additional $SU(3)$ self-dual torus is simply an observer. Using the fiber-base flip of the $\ff_4$ singularity yields a consistent quiver given as  
 \begin{align}
\lbrack \fso_{32}\rbrack     \, \,     {\overset{\mathfrak{sp}_8}{1 }}  \, \,   {\overset{\mathfrak{so }_{16}}{4}}  \, \,  1  \, \,    2 \, \,  \underset{[N_F=1]}{\overset{\mathfrak{su}_2}{2}}   \, \,         \lbrack \fsu_{3 }\rbrack \, , \qquad \vec{l}_{LS}=(1, 1, 3, 2, 1) \, .
\end{align}
Notably, the inserted conformal mater that starts from the $\fso_{16}$ curve looks like that of  $\mathcal{T}(\fe_8,\fsu_3)$. This on the other hand can also be deduced by noting that the $\mathbb{Z}_2$ quotient stops at the $\fso_{16}$ gauge group factor.
 
\paragraph{{\bf LST 3. $\mathbf{E_6^3/\mathbb{Z}_3}$ Flavor: }} 
This case starts to deviate from the usual HW structure in more ways. The underlying singular K3 admits a very nice description in terms of toroidal orbifolds of type $T^2\times T^2/\mathbb{Z}_3$  \cite{Dasgupta:1996ij,Hayashi:2019fsa,Kohl:2021rxy}.
It is easy to see, that the orbifold action fixes the complex structure of both tori to   $\tau = e^{2 \pi i /3}$. One of those tori we identify with the F-theory torus.
As before, the rigid structure only allows for singularities compatible with such a $\tau$ value. In addition there is a $\mathbb{Z}_3$ MW group, which also acts non-trivially on the possible fiber \cite{Dierigl:2020myk} singularity. This non-trivial global $\mathbb{Z}_3$ structure must also act non-trivially on all other gauge groups, which fixes the quiver almost uniquely 
\begin{align}  
\begin{array}{c}
 \lbrack \fe_6\rbrack  \\
 1 \\
 {\overset{\mathfrak{su}_3}{3}} \\
 1 \\
 \lbrack   \fe_6\rbrack \, \, 1 \,\, {\overset{\mathfrak{su}_3}{3}}\,  \,  1 \, {\overset{\mathfrak{e}_6}{6}} \, \, 1 \,\, {\overset{\mathfrak{su}_3}{3}}\,  \,  1 \,  \lbrack \fe_6\rbrack 
\end{array}
 \begin{array}{c} 
\quad \qquad \qquad  1, \\
\quad  \qquad \qquad 1, \\
\quad \qquad \qquad 2, \\
\, , \qquad  \vec{l}_{LS}= ( 1 ,\,\, 1,\,  \,  2, \, 1, \, \, 2, \,\, 1,\,  \,  1 \, ) \, .
\end{array} 
\end{align} 
This quiver also admits a manifest $\mathbb{S}_3$ symmetry permutation symmetry
\paragraph{{\bf LST 4. $\mathbf{ (SO(14)\times SU(12))/\mathbb{Z}_4)  }$ Flavor:}}
Here the $\mathbb{Z}_4$ MW group again restricts the admissible fibers and in particular conformal matter theories to those classified in 
 \cite{Dierigl:2020myk}. From those a matching quiver can be constructed given as 
\begin{align}
\lbrack \fso_{14}\rbrack     \, \,     {\overset{\mathfrak{sp}_2}{1 }}  \, \,   {\overset{\mathfrak{so }_{10}}{4}}  \, \,  1  \, \,    {\overset{\mathfrak{su}_4}{2}} \, \,   {\overset{\mathfrak{su}_8}{2}}  \, \,            \lbrack \fsu_{12}\rbrack \, ,  \qquad \vec{l}_{LS}=(1, 1, 3, 2, 1) \, .
\end{align}

 \paragraph{{\bf 5. $\mathbf{  U(18)/\mathbb{Z}_3   }$ Flavor:}} For this case, we again have restricted monodromies due to the $\mathbb{Z}_3$ MW group. I.e. when adding an $\fsu_3$ type of singularity one obtains the chain
\begin{align}
   \, \,   {\overset{\mathfrak{su }_{3}}{3}}  \, \,  1  \, \,    {\overset{\mathfrak{su}_6}{2}} \, \,   {\overset{\mathfrak{su}_{12}}{2}}  \, \,            \lbrack \mathfrak{u}_{18}   \rbrack  \, ,  \qquad \,  \vec{l}_{LS}=(1, 3, 2, 1) \, .
\end{align}
Note here that the $\mathfrak{u}_1$ flavor factor fits nicely into the full flavor group of the quiver diagram: The BPS string that wraps the $1$ curve is an E-string for which we expect an $E_8$ flavor group. There we find an 
  $\fsu_3 \times \fsu_6$ subgroup of the $ \fe_8$ to be gauged
  by the neighbouring curves
which  leaves a residual $\mathfrak{u}_1$ sub-group. This flavor factor is reflected in the rank one Mordell-Weil group.   
\paragraph{{\bf 6. $\mathbf{ (SO(20)\times E_7)/\mathbb{Z}_2 \times U(1)  }$  Flavor:}}
The final configuration is similar as before, but with an $\mathbb{Z}_2$ MW group as well as an overall $\mathfrak{u}_1$ part  
\cite{Dierigl:2020myk} we obtain 
\begin{align}
\lbrack \fso_{20}\rbrack     \, \,     {\overset{\mathfrak{sp}_4}{1 }}  \, \,   {\overset{\mathfrak{so }_{12}}{4}}  \, \,  1  \, \,    {\overset{\mathfrak{su}_2}{2}} \, \,   {\overset{\mathfrak{so}_7}{3}}  \, \, 
{\overset{\mathfrak{su}_2}{2}} \, \,  
1 \, \, 
\lbrack \fe_{7}\times \mathfrak{u}_1 \rbrack   \, ,   \qquad \vec{l}_{LS}=(1, 1, 3, 2,1,1, 1) \, .
\end{align} 
There are two E-string curves: The right most has a maximal 
$\fe_7 \times \fsu_2$ subgroup of $\fe_8$ gauge by the neighbouring curves. The middle one however admits only a $\fso_{12} \times \fsu_2$ subgroup gauged. Hence we have another residual $\mathfrak{u}_1$ flavor factors which is realized by the free part of the MW group by the geometry.

\section*{Acknowledgements}
We thank Andreas Braun, Jonathan Heckmann, Joseph Minahan, Thorsten Schimannek and Timo Weigand for discussions. The work of MDZ and ML has received funding from the
European Research Council (ERC) under the European Union’s Horizon 2020 research
and innovation program (grant agreement No. 851931). MDZ also acknowledges
support from the Simons Foundation Grant \# 888984 (Simons Collaboration on Global
Categorical Symmetries). PO has received funding from  the NSF CAREER grant PHY-1848089 and startup funding from Northeastern University by Fabian Ruehle.

%PO: I commented out the bib because this yields crashes. 

 
 \appendix
 \section{Toric rays of compact divisors in Example 1} 
 In Table~\ref{tab:CompactE7} we summarize the toric rays that correspond to the compact divisors $\hat \Delta_4$ of the 4d ambient variety used in Example 1 of Section \ref{sec:DiscreteHolonomy}. Details of the analysis of their fibral structure under the three elliptic fibrations are given in \cite{DelZotto:2022xrh}.
\begin{table}[h!]
\footnotesize{
\begin{tabular}{c}
\begin{tabular}{|c|c| }\hline 
\multicolumn{2}{|c|}{$\fe_7$ $i=0\ldots N$ gauge factors }\\ \hline
$\beta_{1,i} $ & $(-1,2,i,\pm 1  )$ \\
$\beta_{2,i} $ &$ (-1, 2,2i,\pm 2  )$ \\ 
$\beta_{3,i} $ & $(-1,2,3i,\pm 3  )$  \\ 
$\beta_{4,i} $ & $(-1,2,4i,\pm 4  )$  \\
$\alpha_{3,i} $ &$ (-1,1,3i,\pm 3  )$ \\    
$\alpha_{2,i} $ & $(0,1,2i,\pm 2  )$  \\ 
$\alpha_{1,i} $ & $(-1,1,i,\pm 1  )$  \\
$\gamma_{2,i} $ &$ (-1,0,i,\pm 2 )$  \\  
\hline   
\end{tabular}  

\\
\begin{tabular}{|c|c|}\hline
\multicolumn{2}{|c|}{ $\mathcal{T}(\fe_7^{i},\fe_7^{i+1})$ $i=0 \ldots N-1$ Conformal matter  }\\ \hline
$s_{\pm 1}$  &$ (-1,2, 1+4i ,4 )  $ \\
$s_{\pm2 }$ &$ (-1,2,1+3i,3),(-1,1,1+3i,3) $  \\
$s_{\pm3,i }$ & $(-1,2, 1+2i ,2 ),(-1,0,1+2i,2),(0,1,1+2i,2),(-1,2,2+2i,4) $  \\
$s_{\pm4, j}$ & $(-1,2, 2+3i ,3  ),(-1,1,2+3i,3)  $ \\
$s_{\pm5 }$ &$ (-1,2,3+4i ,4 )$ \\ \hline   
\multicolumn{2}{|c|}{ 0-th $ \mathcal{T}(\fe_7 ,\fe_7 )$   Conformal matter  }\\ \hline
$s_{\pm 1}$  &$ (-1,2, -3 ,1 )  $ \\
$s_{\pm2 }$ &$ (-1,2,-2,1),(-1,1,-2,1) $  \\
$s_{\pm3,i }$ & $(-1,2, -1 ,1 ),(-1,0,-1,1),(0,1,-1,1),(-1,2,-2,2) $  \\
$s_{\pm4, j}$ & $(-1,2,-1 ,2  ),(-1,1,-1,2)  $ \\
$s_{\pm5 }$ &$ (-1,2,-1 ,3 )$ \\ \hline 
\multicolumn{2}{|c|}{ N-th $ \mathcal{T}(\fe_7 ,\fe_7 )$   Conformal matter  }\\ \hline
$s_{\pm 1}$  &$ (-1,2, 3+N ,1 )  $ \\
$s_{\pm2 }$ &$ (-1,2,2+N,1),(-1,1,2+N,1) $  \\
$s_{\pm3,i }$ & $(-1,2, 1+N ,1 ),(-1,0,1+N,1),(0,1,1+N,1),(-1,2,2+N,2) $  \\
$s_{\pm4, j}$ & $(-1,2,1+2N ,2  ),(-1,1,1+2N,2)  $ \\
$s_{\pm5 }$ &$ (-1,2,1+3N  ,3 )$ \\ \hline
\multicolumn{2}{|c|}{ E-string Conformal Matter  }\\ \hline
$f_{0} $ & $(0,0,0, 1  )$ \\
$f_{M} $ &$ (0, 0,N,1  )$ \\ \hline
\end{tabular} 

\end{tabular} }

 \caption{ \label{tab:CompactE7}Summary of toric vertices  used to resolve the 4d ambient variety used in the threefold of Example 1.}
 \end{table} 
 \bibliographystyle{amsplain}
 % \bibliographystyle{natbib}
\newpage
  \bibliography{main.bib}
\end{document}